\documentclass[aps,prb,twocolumn,amsmath,amssymb,superscriptaddress,floatfix,nofootinbib,longbibliography]{revtex4-2}
\usepackage[utf8]{inputenc}
\usepackage{amsmath}
\usepackage{amsfonts}
\usepackage{amssymb}

\usepackage{graphicx}
\usepackage{bbold}
\usepackage{comment}
\usepackage{breqn}
\usepackage[dvipsnames]{xcolor}

\usepackage{bm}
\usepackage[unicode]{hyperref}
\hypersetup{colorlinks,citecolor=black,linkcolor=blue}
\usepackage[capitalize]{cleveref}

\usepackage{mathrsfs}
\usepackage[scr=boondox,
             cal=dutchcal]  
            {mathalpha}
\usepackage{calligra}

\begin{document}

\title{Synthetic magnetoelectric response of lattice bosonic insulators}

\author{Gautam K. Naik, Michael O. Flynn, Chris R. Laumann}
\affiliation{Department of Physics\char`,{} Boston University\char`,{} 590 Commonwealth Avenue\char`,{} Boston\char`,{} Massachusetts 02215\char`,{} USA}
\date{\today}

\begin{abstract}
    In the absence of parity and time-reversal symmetries,  insulators can exhibit magnetoelectric responses, in which applied magnetic fields induce charge polarization and, conversely, applied electric fields induce magnetization.
    While there is a long history of the study of magnetoelectric response in fermionic insulators, the same for bosonic insulators has been limited. 
    We consider the magnetoelectric response in lattice insulators built out of charged bosonic degrees of freedom and derive a bulk formula for the corresponding linear response tensor.
    The resulting formulae feature several contributions including a Chern-Simons integral over the bands of the  bosonic excitations.
    We construct several minimal microscopic models that illustrate the ingredients required to obtain a sizable bosonic magnetoelectric response.
    Our formalism can be applied to bosonic Mott insulators subject to synthetic gauge fields and/or tilted potentials as well as to the spinon sector in the Coulomb phase of a $U(1)$ quantum spin liquid.
\end{abstract}
\maketitle

\section{Introduction}\label{sec:introduction}

The linear magnetoelectric polarizability of a three-dimensional insulator is captured by the magnetoelectric tensor,
\begin{align}
    \alpha^i_{j}=\frac{\partial P^i}{\partial B^j}= \frac{\partial M_j}{\partial E_i}
    \label{eq:alpha_defn}
\end{align}
where $E$ and $P$ are the electric field and polarization and $B$ and $M$ are the magnetic field and magnetization. 
The second equality is a thermodynamic Maxwell relation, and thus holds at low frequency in quasi-equilibrium.
While the study of $\alpha$ has a long history in magnetic materials \cite{riverashortreviewmagnetoelectric2009, iniguezFirstPrinciplesApproachLatticeMediated2008,xuFirstPrinciplesApproachesMagnetoelectric2024,spaldinAdvancesMagnetoelectricMultiferroics2019}, a theoretical framework for computing the orbital contribution to $\alpha$ in band insulators was only developed relatively recently ~\cite{qiTopologicalFieldTheory2008,essinMagnetoelectricPolarizabilityAxion2009,essinOrbitalMagnetoelectricCoupling2010,malashevichTheoryorbitalmagnetoelectric2010}.
This was motivated by the discovery of fermionic topological insulators, where the orbital contribution is quantized to a non-zero value even when inversion $\mathcal{P}$ and time-reversal $T$ symmetry cause all other contributions to vanish\cite{qiTopologicalInsulatorsSuperconductors2011,sekineAxionElectrodynamicsTopological2021a}.

\begin{figure}[b]
    \centering
    \includegraphics[width=\linewidth]{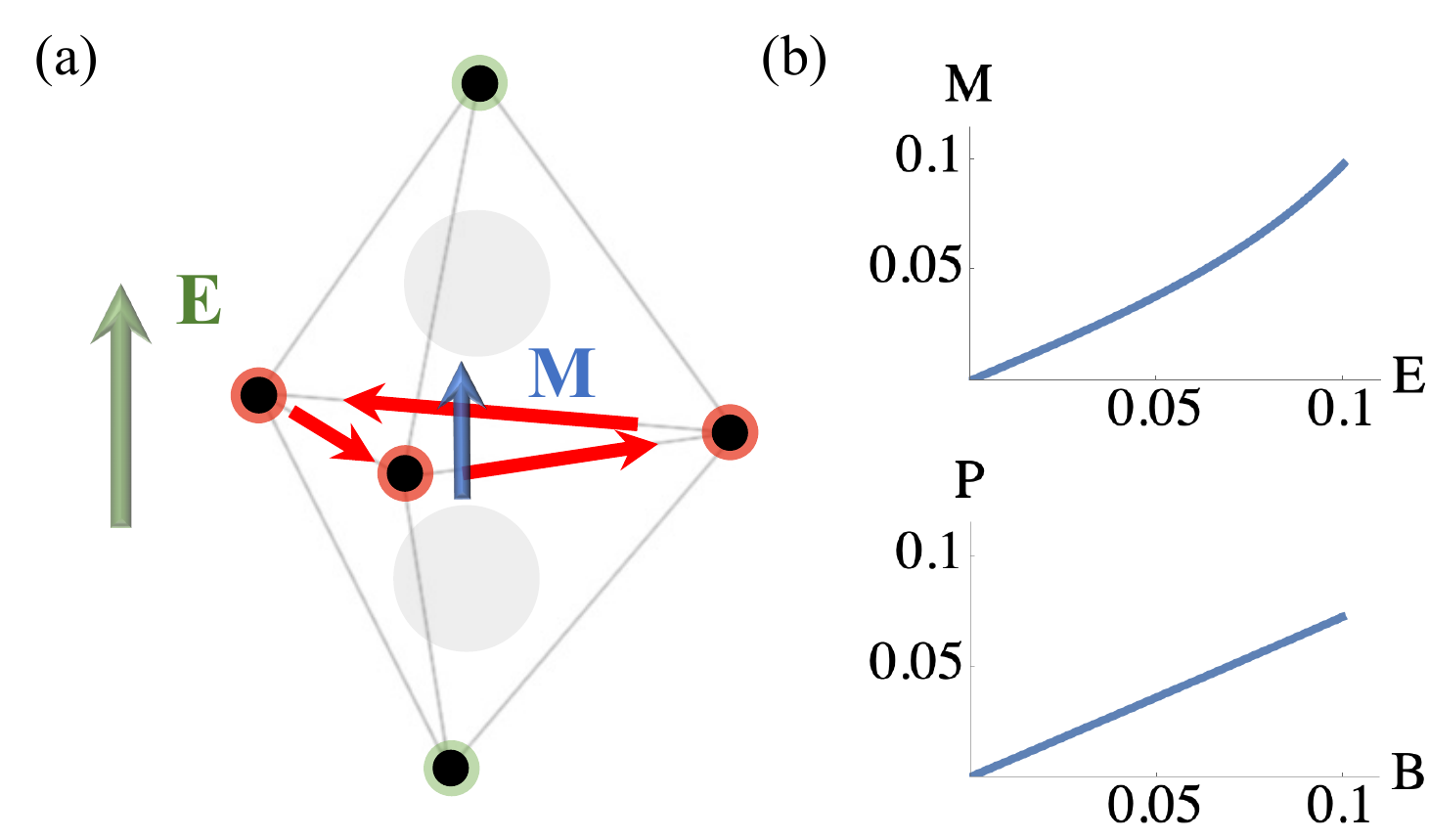}
    \caption{(a) The smallest bosonic model with magnetoelectric response consists of a bipyramid with background monopoles in each tetrahedron and a potential staggered on the red and green sites. 
    Applying an electric field along the three-fold rotation axis leads to a current loop and magnetization. 
    (b, top) Magnetization in response to applied electric field and (bottom) polarization in response to applied magnetic field along the three-fold rotation axis for the bipyramid, with Hamiltonian  \cref{eq:hamiltonian_phi_pi} (parameters  $\lambda=.34,t=1, v=1$ and $m=0.1$). 
    We use units with $\hbar=e=1$. 
    The $x$ and $y$ axes of both plots have dimensions that depend on the length scale, but the zero field slope, which gives the linear magnetoelectric response coefficient, is dimensionless and independent of the length scale.  
    }
    \label{fig:bipyramid_magtization_EM_plots}
\end{figure}

In this article, we consider the magnetoelectric response of non-topological \emph{bosonic} insulators in the absence of inversion and time-reversal symmetry. 
A motivating example is provided by the bosonic Mott insulator in an optical lattice.
Here, the `insulator' blocks transport of the conserved $U(1)$ charge corresponding to that of the underlying neutral atoms. 
These do not couple to a true electromagnetic field; nonetheless,  the magnetoelectric response $\alpha$ can be probed using local potentials and synthetic gauge fields\cite{goldmanLightinducedGaugeFields2014,grossQuantumSimulationsUltracold2017,schaferToolsQuantumSimulation2020}. 
For example, a tilted optical lattice can play the role of an applied electric field, and the magnetization $M_j = \alpha^i_j E_i$ is reflected in lattice scale circulating currents of the bosonic constituents (see \cref{fig:bipyramid_magtization_EM_plots}).

A more exotic physical setting is provided by the $U(1)$ Coulomb quantum spin liquid\cite{udagawaSpinIce2021,henleyCoulombPhaseFrustrated2010,savaryCoulombicQuantumLiquids2012,savaryQuantumSpinIce2016}. 
The spinon sector may be viewed as a bosonic insulator. 
Unlike the atomic Mott insulator, the charge of the spinons couples to an emergent dynamical electromagnetic field. 
Below the spinon gap, the magnetoelectric response $\alpha$ couples into the dynamics of the emergent electromagnetism, appearing like a $\theta$-term in the effective theory.

\begin{figure}
    \centering
    \includegraphics[width=1\linewidth]{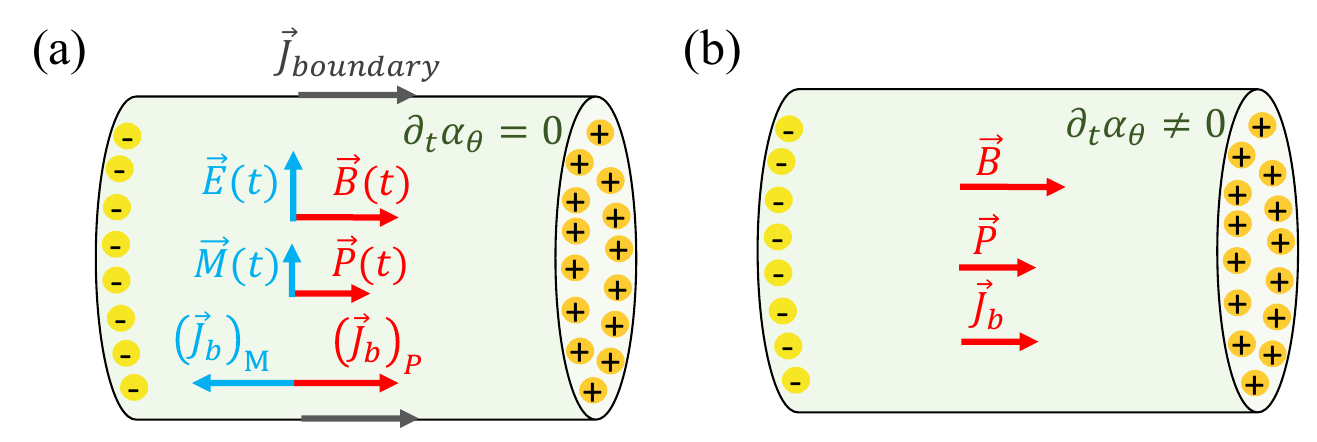}
    \caption{The bulk current within the insulator, $\mathbf{J}_b= \partial_t \mathbf{P} + \mathbf{\nabla} \times \mathbf{M}$, has contributions from the bulk polarization and magnetization.  
    (a) A system with uniform time-independent $\alpha_\theta$ and zero $\tilde\alpha$. 
    Slowly turning on a magnetic field induces time-varying electric fields according to Faraday's law. 
    The magnetoelectric response to these fields leads to cancellation of the bulk current.
    The current that generates the polarization flows only along the boundary surface. 
    (b) In a system with time-varying $\alpha_\theta$ and constant magnetic field, there is no induced electric field, and there is a non-zero bulk current~\cite{essinOrbitalMagnetoelectricCoupling2010}. 
    }
    \label{fig:bulk_current}
\end{figure}

In this article, we derive closed formulae for the magnetoelectric response $\alpha$ of lattice systems of gapped bosonic oscillators with a $U(1)$ charge.
Our formalism applies to the quadratic approximation to the excitations around a mean-field insulating state, and we expect it to be quantitatively well-controlled by the gap of the insulator.
The derivation of a bulk formula for $\alpha$ is more complicated than one might expect due to the Maxwell relation in \cref{eq:alpha_defn}.
More precisely, the magnetoelectric tensor can be decomposed into a pseudo-scalar and a traceless part:
\begin{align}
    \label{eq:alpha_tracedecomp}
    \alpha^i_j = \tilde{\alpha}^i_j + \alpha_\theta \delta^i_j
\end{align}
The problem is that Faraday's law, $\partial_t \bm{B}+\nabla \times \bm{E}=0$, in conjunction with the Maxwell relation, ensures that the pseudo-scalar part $\alpha_\theta$ cancels out of the bulk current response to applied low-frequency fields (see \cref{fig:bulk_current}).
Hence, standard bulk calculations that neglect surface currents fail to compute $\alpha_\theta$.

Despite this difficulty, several approaches have been developed to obtain $\alpha_\theta$ for fermionic systems. 
Ref.~\onlinecite{qiTopologicalFieldTheory2008} utilized a field-theoretic dimensional reduction approach which applies directly only to systems with sufficient symmetry.
They found that $\alpha^i_j = \alpha_{\text{CS}} \delta^i_j$, where $\alpha_{\text{CS}}$ is given by a Chern-Simons integral in momentum space. 
More generally, Essin et al.~\cite{essinOrbitalMagnetoelectricCoupling2010} developed an elegant approach to computing $\alpha$ in general fermionic band insulators by considering an adiabatic protocol in which the bulk Hamiltonian varies in time in the presence of constant magnetic field. 
They found an additional `cross-gap' contribution $\alpha_{\text{G}}$, 
\begin{align}
    \label{eq:alpha_cs_decomposition}
    \alpha^i_j = \alpha_{\text{CS}} \delta^i_j + (\alpha_{\text{G}})^i_j
\end{align}
which, notably, also contributes to the trace $\alpha_\theta$.
We will review this approach in more detail below, as our bosonic derivation mirrors it.
We note that the same result for fermionic insulators was also obtained by considering a constant background electric field in Ref.~\onlinecite{malashevichTheoryorbitalmagnetoelectric2010}. 
There have been a few alternate derivations \cite{chenUnifiedFormalismCalculating2011,mahonMagnetoelectricPolarizabilityMicroscopic2020} as well as generalizations to disordered systems\cite{leungNoncommutativeFormulaIsotropic2013}, interacting systems\cite{everschor-sitteInteractionCorrectionMagnetoelectric2015,schmeltzerMagnetoelectricEffectInduced2013} and response at finite frequency \cite{mahonMagnetoelectricResponseOptical2020,duffMagnetoelectricPolarizabilityOptical2022,shinadaQuantumTheoryIntrinsic2023}, when the Maxwell relation in \cref{eq:alpha_defn} does not hold.

\begin{figure}
    \centering
    \includegraphics[width=\linewidth]{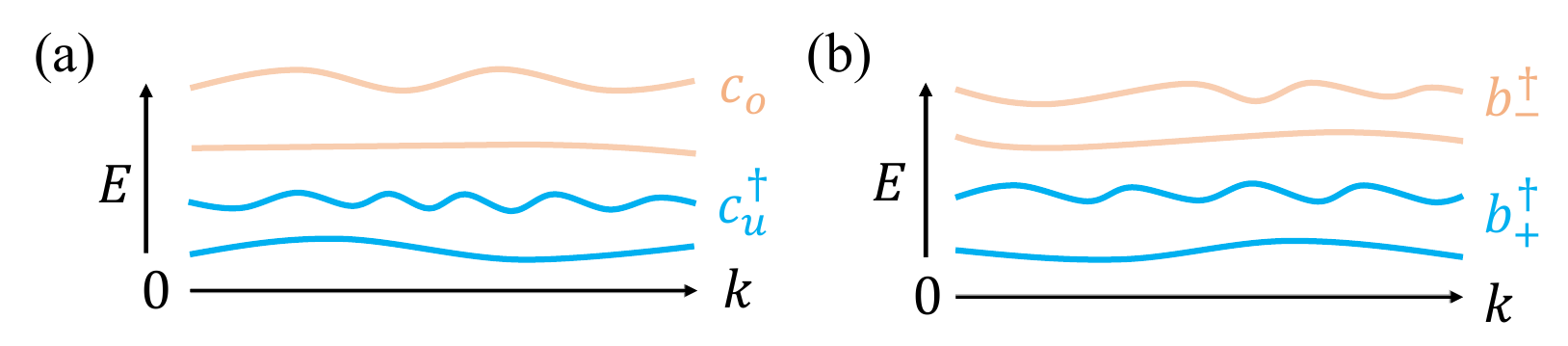}
    \caption{Band structure of fermionic and bosonic insulators. (a) The single particle excitations of fermionic systems are obtained by adding electrons to unoccupied states (blue) or removing electrons from occupied states (orange). All of these electron and hole modes have positive energy.  (b) In bosonic insulators, there are excitation bands corresponding to the positive and negative charged mode creation operators. Stable insulators require a finite positive neutral excitation gap. 
    In such systems, the chemical potential can be adjusted  so that all of the charged modes have positive excitation energy.
    }
    \label{fig:band_structure}
\end{figure}

Our main result is the following closed formula for the magnetoelectric tensor expressed as a trace over the bosonic excitation modes in momentum space ($e=\hbar=1$),
\begin{widetext}
\begin{align}
\alpha_{\text{CS}}
&= \frac{1}{8}\int_0^1 d\beta \int_{\text{BZ}} \frac{d^3{k}}{(2\pi)^3} \epsilon_{\mu\nu\gamma\lambda} \ \text{Tr}\ P \left( \partial^\mu P \partial^\nu P - \partial^\nu P \partial^\mu P \right) P \left( \partial^\gamma P \partial^\lambda P - \partial^\lambda P \partial^\gamma P \right)
\label{eq:alpha_cs}\\
(\alpha_{\text{G}})^i_j &= \int_{\text{BZ}} \frac{d^3{k}}{(2\pi)^3} \sum_{n, m} \frac{ \text{Re} \ \text{Tr} \  P_{-n} \ \partial^i P \ \epsilon_{j\mu\nu} \ P_{+m} \{\partial^\mu {h_o}, \partial^\nu P \}  + 2\  \text{Im} \ \text{Tr} \ P_{-n} \  \partial^i P P_{+m} (\partial h'/\partial B^j) }{E_{-n} + E_{+m}}
\label{eq:alpha_G}
\end{align}
\end{widetext}
As promised, $\beta$ is an adiabatic parameter relating our Hamiltonian of interest to a reference Hamiltonian with vanishing~\footnote{More generally, Eq.~\eqref{eq:alpha_cs} can be interpreted as the change in $\alpha_{\text{CS}}$ on tuning $\beta$.} $\alpha$. 
The projector $P$ picks out the annihilation operators for the negatively charged modes; this plays a role analogous to projection onto occupied states in the fermionic case, see \cref{fig:band_structure}. 
The `cross-gap' term depends on the energies $E_{\pm m}(\bm{k})$ of the positively and negatively charged bosonic bands and their corresponding mode projectors $P_{\pm m}(\bm{k})$.
The Hamiltonian appears explicitly in $\alpha_{\text{G}}$ through the dynamical matrix $h = h_o + h'$, which governs the mode dynamics. 
Precise mathematical definitions can be found in \cref{sec:formalism}.

The `Chern-Simons' contribution $\alpha_{\text{CS}}$ in  \Cref{eq:alpha_cs} can be rewritten in terms of a second Chern form: 
\begin{align}
    \alpha_{\text{CS}} = -\frac{1}{8}\int_0^1 d\beta \int_{\text{BZ}} \frac{d^3{k}}{(2\pi)^3} \epsilon_{\mu\nu\gamma\lambda} \ \text{Tr}\ F^{\mu\nu}F^{\gamma\lambda}
    \label{eq:alpha_cs_chernform}
\end{align}
where, $F^{\mu\nu}= i P\left(\partial^\mu P \partial^\nu P - \partial^\nu P \partial^\mu P \right) P$. 
This 4D second Chern form can further be rewritten as a 3D momentum space Chern-Simons form of the corresponding Berry connection along the boundary of the $\beta$ integral. 
However, this way of rewriting the term and ignoring the adiabatic change from a reference Hamiltonian introduces a gauge dependence to the integral. 
The gauge freedom also makes this integral harder to compute numerically.
Hence, in this article, we only use the second Chern form to compute $\alpha_{\text{CS}}$.

The expressions we have in \cref{eq:alpha_cs,eq:alpha_G} are similar to those for fermionic insulators~\cite{essinOrbitalMagnetoelectricCoupling2010}, with the similarity most evident in the electron-hole picture (see \cref{fig:band_structure}) of the insulator.
The distinction is that the bosonic mode operators are obtained by generalized Bogoliubov transformations, while the fermionic modes are obtained by unitary transformations.
This has several consequences; for example, the projector $P$ is self-adjoint with respect to a conjugate symmetric sesquilinear form with mixed signature rather than a more familiar positive definite inner product. 

Topological insulators show a quantized non-zero isotropic magnetoelectric response without breaking $\mathcal{P}$ or $T$ symmetry.
There are known non-interacting fermionic topological insulators, and in these systems, the Chern-Simons integral gives the corresponding quantized non-zero value for $\alpha_{\text{CS}}$~\cite{essinMagnetoelectricPolarizabilityAxion2009}.
However, to the best of our knowledge, all known examples of bosonic topological insulators are strongly interacting~\cite{vishwanathPhysicsThreeDimensionalBosonic2013,levinExactlySolubleModels2011,swingleCorrelatedTopologicalInsulators2011}. 
We believe our formalism applies to $\mathcal{P}$ and $T$ breaking Hamiltonians that are effectively quadratic and can be obtained by topologically deforming a trivial Hamiltonian.
We leave the question of calculating this response for bosonic topological insulators and symmetry breaking around such systems as an open question.

Note that there are several existing bosonic band formulations and mappings of bosonic systems to fermionic systems \cite{shindouTopologicalChiralMagnonic2013,zhangTopologicalMagnonInsulator2013,matsumotoThermalHallEffect2014,iacoccaTopologicallyNontrivialMagnon2017,luMagnonBandTopology2018a,mcclartyTopologicalMagnonsReview2022}. 
These existing mappings are usually used to simplify the calculation of topological properties of the bosonic bands. The mapping we introduce simply points out the mathematical resemblance between the calculations done in $U(1)$ conserving bosonic systems and $U(1)$ conserving fermionic systems and allows us to write out results that apply to bosonic systems from existing results that apply to fermionic systems.

Our formalism applies to generic quadratic lattice systems of bosonic oscillators with a globally conserved $U(1)$ charge. 
This charge may be understood as the `angular momentum' of the 2D harmonic oscillator at each lattice site. 
We describe these oscillators in terms of a pair of complex scalars, $\Phi_{\bm{r}}$ and $\Pi_{\bm{r}}$, at each lattice site $\bm{r}$, with canonical commutator relations $[\Phi_{\bm{r}},\Pi_{\bm{r}'}^\dagger ]= i \delta_{\bm{r}\bm{r}'}$. 
The general Hamiltonian can be expressed,
\begin{align}
    \mathcal{H} = 
    \sum_{\bm{rr'}}\begin{pmatrix} \Pi^\dagger_{\bm{r}}&\Phi^\dagger_{\bm{r}} \end{pmatrix}
    \begin{pmatrix} M^{-1}_{\bm{rr'}} & iV_{\bm{rr'}} \\
    -i V^\dagger_{\bm{rr'}} & K_{\bm{rr'}} \end{pmatrix}
    \begin{pmatrix} \Pi_{\bm{r'}} \\
    \Phi_{\bm{r'}} \end{pmatrix}
    \label{eq:hamiltonian_phi_pi}
\end{align}
where we view $M$ as a mass matrix and $K$ as a `spring' coupling matrix.
The off-diagonal $V$ matrix may be viewed as a generalized potential, as the diagonal part couples to the local charge, $Q_{\bm{r}}$.

There are many experimental realizations of fermionic magnetoelectric response in multiferroics\cite{spaldinAdvancesMagnetoelectricMultiferroics2019,xuFirstPrinciplesApproachesMagnetoelectric2024,liTopologicalInsulatorsSemimetals2021}. However, materials in which the orbital contribution to the response is dominant are primarily, but not limited to, topological insulators\cite{okadaTerahertzSpectroscopyFaraday2016,dziomObservationUniversalMagnetoelectric2017,wuQuantizedFaradayKerr2016,sekineAxionElectrodynamicsTopological2021a,malashevichFullMagnetoelectricResponse2012,gobelMagnetoelectricEffectOrbital2019,kammermeierInplaneMagnetoelectricResponse2019}.

The rest of the article is organized as follows: \Cref{sec:toy_model} introduces simple bosonic models that show a magnetoelectric response. 
\Cref{sec:formalism} introduces the general formalism we use to study bosonic systems with a global U(1) symmetry. 
In \cref{sec:numerics}, we present numerical results for a lattice bosonic insulator. We conclude and discuss future prospects in \cref{sec:outlook}.
The appendix includes the following: 
\cref{sec:toy_model_hamiltonian} provides details to the explicit form of the Hamiltonians we use.
\cref{sec:properties_of_the_correlation_matrix} elucidates the proof for two of the key properties of the correlation matrices introduced in \cref{sec:formalism}. In \Cref{sec:derivation-of-the-magnetoelectric-tensor}, we go over the derivation of the expressions in \cref{eq:alpha_cs,eq:alpha_G}. \Cref{sec:group_structure_of_the_generalized_Bogoliubov_transformation} clarifies the group structure of the diagonalization of Hamiltonians in \cref{eq:hamiltonian_phi_pi}. \Cref{sec:generality_of_formalism} shows how the formalism we introduce in \cref{sec:formalism} can be generalised to any $U(1)$ symmetric quadratic bosonic Hamiltonian.

\section{Toy Model}\label{sec:toy_model}

In this section, we illustrate the magnetoelectric effect by introducing small models that break all the required symmetries to have a non-zero $\alpha$. These `toy models' help illustrate that additional microscopic ingredients are required to observe a magnetoelectric response in bosonic systems rather than fermionic systems. We also use the toy models to benchmark our numerical computation of the integrals in \cref{eq:alpha_cs,eq:alpha_G}.

\begin{figure}[t!]
    \centering
    \includegraphics[width=.95\linewidth]{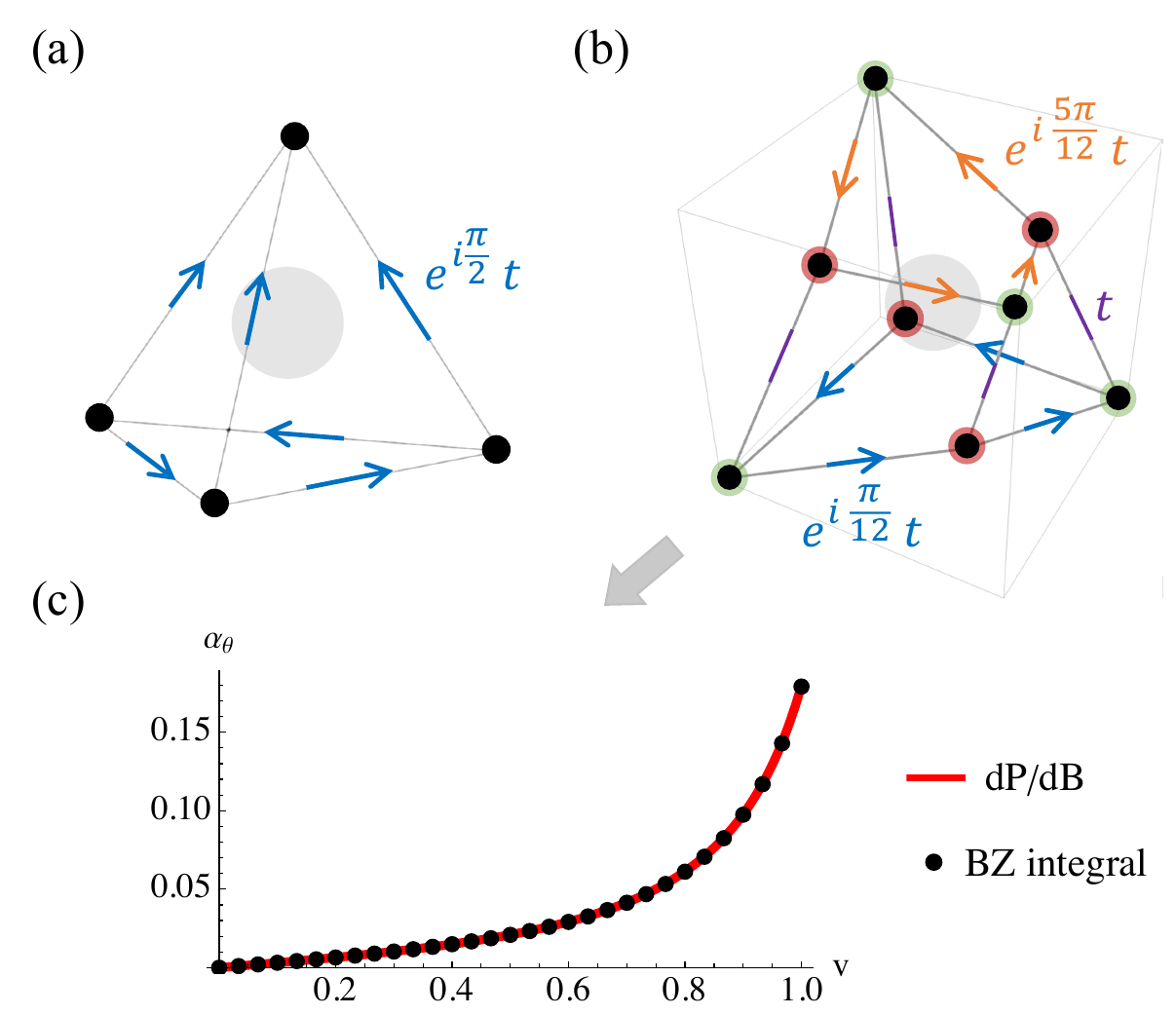}
    \caption{Minimal systems for isotropic magnetoelectric response. 
    (a) For quadratic fermions, a single tetrahedron with an enclosed background magnetic monopole exhibits a magneto-electric response.  
    Blue arrows indicate a gauge in which the oriented hop carries phase $e^{i\pi/2}$.
    (b) For quadratic bosons, a minimal isotropic model is that of a deformed cube with a magnetic monopole and an alternating potential. 
    Colored arrows indicate oriented hops with amplitude given in the figure. 
    The green (red) sites have positive (negative) local potential. 
    (c) The isotropic part of the magnetoelectric tensor $\alpha_\theta$ (where $e=\hbar =1$) as a function of the charge staggering parameter $v$ for the deformed cube model with $\lambda=2.6,t=1$, $m=0.1$ and the inner vertices pushed in halfway to the center of the cube. 
    The red curve is obtained from the zero field slope of the $P$ vs $B$ curve of a single deformed cube. 
    The black dots are the points obtained by performing the integrals in \cref{eq:alpha_cs,eq:alpha_G}. 
    Note that the insulator gap decreases with increasing $v$. }
    \label{fig:toy_models_schematic_&_plot}
\end{figure}

Before turning to bosonic toy models, let us review the simplest \emph{fermionic} hopping model~\cite{essinOrbitalMagnetoelectricCoupling2010} exhibiting a magnetoelectric response.
This consists of fermions hopping on a tetrahedron containing a background magnetic monopole. 
The Hamiltonian is $\mathcal{H}_f= -\sum_{\langle {\bm{r}},{\bm{r}}'\rangle} c_{\bm{r}}^\dagger t_{{\bm{r}}{\bm{r}}'}c_{{\bm{r}}'}$, where, $t_{{\bm{r}}{\bm{r}}'}$ is chosen such that the Aharonov-Bohm phase from hopping around any face of the tetrahedron is $\pi/2$ (as shown in \cref{fig:toy_models_schematic_&_plot}(a)). 
It is straightforward to diagonalize this model and derive the isotropic magneto-electric response, $\alpha^i_j = \frac{1}{\sqrt{6}} \frac{e^2}{\hbar} \delta^i_j$, as shown in Ref.~\onlinecite{essinOrbitalMagnetoelectricCoupling2010}.

It is instructive to review the symmetries of the fermionic monopole-tetrahedron to see how they permit a non-zero $\alpha$.
The system is symmetric under the proper rotational symmetries of the tetrahedron, which leaves both the magnetic monopole field and the tetrahedron invariant.
This ensures that the magnetoelectric response $\alpha^i_j = \alpha_\theta \delta^i_j$ is isotropic.
The magnetic monopole breaks the improper reflection symmetries, $M_i$, of the tetrahedron as the monopole field reverses under such reflections. 
However, as time-reversal $T:c\to c, i\to -i$ also reverses the monopole field, $H_f$ is symmetric under the combined action of $M_i T$. 
Finally, $H_f$ is symmetric under (unitary) charge conjugation, $C: c_{\bm{r}} \to c^\dagger_{\bm{r}}$. 
Mathematically, all of the symmetries must be implemented along with appropriate gauge transformations to leave $\mathcal{H}_f$ invariant.
Crucially, these symmetries are all \emph{proper} in the 3+1D sense, and thus, non-zero $\alpha$ is permitted.

Let us now attempt to construct a bosonic system with a magnetoelectric response.
The simplest attempt is to consider bosonic modes, $a_{\bm{r}}$, at every corner of the tetrahedron with hopping the same as that for the fermionic toy model, i.e. $\mathcal{H}_b=-\sum_{\langle {\bm{r}},{\bm{r}}'\rangle} a_{\bm{r}}^\dagger t_{{\bm{r}}{\bm{r}}'}a_{{\bm{r}}'}$. However, this model has bosonic negative energy modes and is unstable. 
If we add diagonal terms of the form $\lambda \sum_{\bm{r}}a_{\bm{r}}^\dagger a_{{\bm{r}}}$, and make $\lambda$ large enough to ensure a finite positive charge gap in the system, we obtain a rather trivial insulator.
The ground state is the Fock vacuum of the $a$-modes, which is unperturbed by any perturbations to $\mathcal{H}_b$.

To obtain a more interesting insulator, we must allow both positively and negatively charged excitations.
We consider two bosonic modes, $a_{+\bm{r}}$ and $a_{-\bm{r}}$, at every site and set the local charge to be $Q_{\bm{r}}=a^\dagger_{+\bm{r}} a_{+\bm{r}} -a^\dagger_{-\bm{r}} a_{-\bm{r}}$. With this setup, we obtain a non-trivial ground state by coupling the charges via terms of the form $(a^\dagger_{+\bm{r}}a^\dagger_{-\bm{r}'}+a_{+\bm{r}}a_{-\bm{r}'})$. 

A perhaps more natural way to describe a pair of bosonic modes at each site is with complex scalars, $\Phi_{\bm{r}}$ and $\Pi_{\bm{r}}$, which satisfy the commutation relations $[\Phi_{\bm{r}},\Pi_{\bm{r}'}^\dagger ]= i \delta_{\bm{r}\bm{r}'}$. The local charge is the angular momentum of the 2D oscillator, $Q_{\bm{r}}= i\left( \Pi^\dagger_{\bm{r}}\Phi_{\bm{r}} - \Pi_{\bm{r}} \Phi^\dagger_r\right)$.
We now consider toy models with a Hamiltonian of the form
\begin{align}
    \mathcal{H} = 
    \sum_{\bm{rr'}}\begin{pmatrix} \Pi^\dagger_{\bm{r}}&\Phi^\dagger_{\bm{r}} \end{pmatrix}
    \begin{pmatrix} m^{-1}\delta_{\bm{r}\bm{r'}} & i v V_{\bm{r}}\delta_{\bm{r}\bm{r'}} \\
    -iv V_{\bm{r}}\delta_{\bm{r}\bm{r'}} & \lambda \delta_{\bm{r}\bm{r'}} + t K_{\bm{rr'}} \end{pmatrix}
    \begin{pmatrix} \Pi_{\bm{r'}} \\
    \Phi_{\bm{r'}} \end{pmatrix},
    \label{eq:simple_hamiltonian_pi_phi}
\end{align}
where, $m,t, \lambda$ and $v$ are real parameters.
Here, $m$ is the uniform local mass of each oscillator, $t$ determines the strength of the off-diagonal nearest neighbor coupling matrix $K$, $\lambda$ gives the diagonal couplings, and $v$ determines the strength of a staggered site-local charge potential, $V_{\bm{r}} \in \pm 1$. 
For any choice of geometry (encoded in $K$ and $V_r$), this produces a three-dimensional phase space as the overall scale of $\mathcal{H}$ is unimportant for the dimensionless magnetoelectric response $\alpha$.

Let us now consider a bosonic tetrahedron-monopole system with complex scalars attached to each corner. 
We represent the Hamiltonian for this system as $\mathcal{H}_{TM}$ and it is given by \cref{eq:simple_hamiltonian_pi_phi} with the coupling matrix $K_{\bm{rr'}}$ matching that of the fermionic hopping matrix $t_{\bm{rr'}}$.
One might expect that this model has all the ingredients required to exhibit non-zero $\alpha$. However, it turns out that the symmetries of this bosonic system aren't the same as that of the corresponding fermionic system. 
The magnetic monopole still breaks time reversal and mirror symmetries while leaving the proper rotational symmetries of the tetrahedron intact.
However, the charge conjugation symmetry $C$ behaves quite differently. 
For the bosonic system, time reversal is defined by $T: \ \phi_{\bm{r}} \rightarrow \phi_{\bm{r}}, \ \Pi_{\bm{r}} \rightarrow -\Pi_{\bm{r}} , \ i \rightarrow -i $ and charge conjugation is defined by $C:\ \phi_{\bm{r}} \rightarrow \phi^\dagger_{\bm{r}}, \ \Pi_{\bm{r}} \rightarrow \Pi^\dagger_{\bm{r}}$. One can check that the $\mathcal{H}_{TM}$ model has $CT$ symmetry.
$\alpha$ is odd under the action of $CT$, and this implies that the bosonic model will not show a magnetoelectric response unless $CT$ is broken.

To break $CT$, we can add non-uniform charge potentials that explicitly break all charge conjugation symmetries. One way to do this while still maintaining some rotational symmetries in 3D is to consider a bipyramid with monopoles in both tetrahedrons and opposing potentials at the apexes and the base (see \cref{fig:bipyramid_magtization_EM_plots}). 
Although this five-site bipyramid model is the simplest \footnote{More precisely, this is the smallest model of the form in \cref{eq:simple_hamiltonian_pi_phi} with nonzero $\alpha$ and sufficient symmetry to rule out a dipole moment and magnetization in the absence of applied fields.} bosonic model that shows a magnetoelectric response, it is anisotropic.
\cref{fig:bipyramid_magtization_EM_plots} shows the response of this model along the 3-fold symmetry axis.

A simple model with isotropic magnetoelectric response has the geometry of a cube deformed such that four of the eight corners are pushed in (as shown in \cref{fig:toy_models_schematic_&_plot}(b)). 
The Hamiltonian for this system is given by \cref{eq:simple_hamiltonian_pi_phi}, with the phases in the spring coupling matrix elements $K_{{\bm{r}}{\bm{r}}'}$ chosen so that the deformed cube encloses a negative monopole, and $V_r$ chosen to be positive at the outer corners and negative at the inner corners. The only remaining symmetries of this system are the proper tetrahedral rotations, which ensure that the system has an isotropic response.

Tiling copies of this toy model along a cubic lattice gives us a system that has translational symmetry. For such a system of non-interacting deformed cubes, the polarization and hence $\alpha$ is the same as that of the single deformed cube up to a geometrical factor. 
The polarization of a single deformed cube at small applied magnetic fields can be computed by diagonalizing a small matrix (of size $16\times16$).
Hence, such a system forms a simple toy example to test the validity of the expressions in \cref{eq:alpha_cs,eq:alpha_G}. 
We use numerical integration of these expressions to compute $\alpha$ and show that this matches with what is obtained from our numerics of a single deformed cube in  \cref{fig:toy_models_schematic_&_plot}(c). Breaking $CT$ by a non-uniform $V_r$ is a generic way to tune from a model with $\alpha=0$ to $\alpha\neq 0$. So, we use the strength $v$ of such fields as the adiabatic parameter $\beta$ in \cref{eq:alpha_cs}.

\section{Analyzing Quadratic Bosonic Systems with a Conserved Charge}\label{sec:formalism}

\begin{table}
\begin{tabular}{|l|l|}
\hline
\multicolumn{1}{|c|}{\textbf{Fermions}}& \multicolumn{1}{c|}{\textbf{Bosons}}\\
\hline
$H$ :  Hamiltonian matrix&$h$ : Dynamical matrix\\
$\rho$ : Density Matrix&$C$ : Correlation matrix\\
$E$ : Energy&$\Lambda$ : Frequency\\
$U$ : Unitary &$R$ : Generalized Bogoliubov \\
\qquad diagonalization& \qquad diagonalization (\cref{sec:group_structure_of_the_generalized_Bogoliubov_transformation})\\
\hline
\end{tabular}
\caption{ Summary of the replacements required to translate the existing fermionic derivation\cite{essinOrbitalMagnetoelectricCoupling2010} to one that applies for bosonic systems. For a fermionic system characterized by the  Hamiltonian $\mathcal{H}_f= c^\dagger H c$, all observables can be computed from the single body density matrix $\rho_{ij}=\langle c^\dagger_i c_j \rangle$. \cref{eq:c^2=c,eq:time_ev_C} are key examples that show that $C$ and $h$ can be replaced by $\rho$ and $H$, respectively, to translate between expressions that hold for bosonic and fermionic systems.}
\label{tab:map_fermions_bosons}
\end{table}

After identifying the correct mathematical definitions to translate from free fermions to general $U(1)$ conserving quadratic bosons, our derivation of \cref{eq:alpha_cs,eq:alpha_G} algebraically mirrors that of Essin et al\cite{essinOrbitalMagnetoelectricCoupling2010}. 
In this section, we present the appropriate mathematical dictionary (see \cref{tab:map_fermions_bosons}) by introducing the formalism to analyze, solve and compute ground state observables within quadratic bosonic systems  with a Hamiltonian of the form in \cref{eq:hamiltonian_phi_pi}.
While the properties of the relevant mathematical objects are somewhat different,  particularly in that the objects are self-adjoint under different sesquilinear forms, ultimately the derivation goes through. 
For completeness, we include the bosonic derivation in \cref{sec:derivation-of-the-magnetoelectric-tensor}.
The key definitions and properties required to push through the derivation as well as make sense of the terms in the resulting \cref{eq:alpha_cs,eq:alpha_G} are presented below.

In quadratic fermionic systems, unitary diagonalization of the Hamiltonian matrix that accompanies the fermionic modes allows one to obtain all fermionic eigenmodes.
In quadratic bosonic systems, unitary diagonalization of the Hamiltonian matrix accompanying the bosonic modes does not give bosonic operators. Instead, a generalized Bogoliubov diagonalization of the dynamical matrix, i.e. the matrix that gives the equations of motion (EOM) of the bosonic modes, is required to obtain the eigenmodes \cite{colpaDiagonalizationQuadraticBoson1978,kusturaQuadraticQuantumHamiltonians2019a}.

\paragraph{Spinor Formalism---}
We consider a generic lattice bosonic system with a $U(1)$ symmetry and $N$ sites, each of which can be excited with both positively and negatively charged excitations.
We introduce a Nambu spinor, $\psi=\left[\begin{smallmatrix}  \Pi\\  \Phi \end{smallmatrix}\right]$, with $\Phi$ and $\Pi$ being a complex scalar field and its conjugate momentum, so that $[\Phi,\Pi^\dagger] =i\mathbb{1}_{N\times N}$. 
In terms of the spinor $\psi$, the canonical commutation relations can be expressed 
\begin{align}
    [\psi,\psi^\dagger]=\sigma^y,
\end{align}
where, the right-hand side is understood to be the $2N\times 2N$ Pauli matrix $\sigma^y\otimes \mathbb{1}_{N\times N}$.
The most general quadratic Hamiltonian $\mathcal{H}$ which respects the $U(1)$ symmetry of the complex scalar can be written as
\begin{align}
    \mathcal{H}= \psi^\dagger \sigma^y h \ \psi \quad \text{ with EOM } \quad i\partial_t\psi = h \psi,
    \label{eq:H_with_psi}
\end{align} 
where $h$ is the $2N \times 2N$ dynamical matrix governing the equations of motion for the modes.
The conserved charge is,
\begin{align}
    Q= i(\Pi^\dagger \Phi - \Pi \Phi^\dagger)=- \psi^\dagger \sigma^y \psi -N.
    \label{eq:total_charge}
\end{align}

\paragraph{Self-Adjointness and the Sequilinear Form---}
Hermiticity of $\mathcal{H}$ requires that $\sigma_y h$ is Hermitian as a matrix, while the dynamical matrix $h$ need not be. 
However, $h$ can be seen as a linear map acting on the space of all charge-increasing mode operators ($u= u^i \psi_i,u^i \in \mathbb{C}$).
This linear map is self-adjoint with respect to the non-degenerate sesquilinear form:
\begin{align}
    (u,v) = [u^\dagger,v] = (u^i)^* [\psi_i^\dagger,\psi_j] v^j = - (u^i)^* \sigma^y_{ij} v^j 
    \label{eq:bilinear_form}
\end{align}
This form is conjugate symmetric and can be viewed as a complex inner product with mixed signature $(N,N)$. 

\paragraph{Bogoliubov Diagonalization---}
The system is diagonalized by finding the similarity transformation, $R$, that diagonalizes the EOM in \cref{eq:H_with_psi}:
\begin{align}
    B=R\psi=\left[\begin{smallmatrix}  b_{-}\\  b_{+}^\dagger \end{smallmatrix}\right] \quad \text{and}\quad
    i\partial_t B=\Lambda B 
\end{align}
where, $\Lambda$ is a diagonal frequency matrix and $B$ is a new set of bosonic operators that contains $N$ annihilation operators, $b_-$, of negatively charged modes and $N$ creation operators, $b_+^\dagger$, of positively charged modes. These operators satisfy the commutation relations
\begin{align}
    [B,B^\dagger]= \sigma^z.
\end{align} 
The positive (negative) charged modes are the modes whose creation operators $b_+^\dagger$ ($b_-^\dagger$) increase (decrease) the total charge of a state by one.
The required similarity transformation, $R$, that diagonalizes the dynamical matrix and gives operators with bosonic commutation relations should satisfy:
\begin{align}
     h=R^{-1} \Lambda \ R \quad \text{and} \quad   R\  \sigma^y R^\dagger = \sigma^z,
\label{eq:h_diag_R}
\end{align}
where, $\Lambda$ is a diagonal matrix with the frequencies of the bosonic modes.
The diagonalized Hamiltonian can be expressed as:
\begin{align}
    \mathcal{H}= B^\dagger \sigma^z \Lambda \ B,   
\label{eq:H_daig_B}
\end{align}
with the energy spectrum of the bosonic modes given by the diagonal matrix $E=\sigma^z \Lambda$.

As $h$ is not hermitian, the right and left eigenvectors of $h$ are not related by a complex conjugate transpose. Nonetheless, we can express the dynamical matrix as 
\begin{align}
&h=\sum_n \Lambda_{nn} \mathcal{w}_n \mathcal{v}^\dagger_n 
\end{align}
where, $\mathcal{w}_n$ ($\mathcal{v}^\dagger_n$) are column (row) vectors that are right (left) eigenvectors of the matrix $h$. 
The eigenvectors form a basis and satisfy conditions
\begin{align} 
\mathcal{v}^\dagger_m \cdot \mathcal{w}_n = \delta_{mn} \quad \text{and} \quad \sum_n \mathcal{w}_n \mathcal{v}^\dagger_n= \mathbb 1.
\end{align}
The diagonalization allows us to construct projectors on to the  $n^{\text{th}}$ bosonic mode with positive and negative charge,
\begin{align}
    P_{-n} &= \mathcal{w}_n\mathcal{v}^\dagger_n=R^{-1} \Gamma_n R \nonumber\\
    P_{+n} &= \mathcal{w}_{N+n}\mathcal{v}^\dagger_{N+n}=R^{-1} \Gamma_{N+n} R,
    \label{eq:projector_nth_mode}
\end{align} 
where, $\Gamma_n$ is a matrix with only a single non-zero element viz. the $n^{\text{th}}$ element along the diagonal being one. The above equation gives the matrix representation of these projectors in the $\psi$ operator basis and these matrices are not Hermitian. However, similar to $h$, these projectors viewed as linear maps on the operator space are self-adjoint under the sesquilinear form in \cref{eq:bilinear_form}.

\paragraph{Ground State Correlations---}
In the ground state of quadratic systems, all observables follow from bilinear correlators (by bosonic Wick's theorem).  
We define the correlation matrix $C$ ($C_B$) of a state in the $\psi$ ($B$) operator basis to be the following expectation value:
\begin{align}
    C= \langle \psi \psi^\dagger \rangle \sigma^y \quad \text{and} \quad C_B = \langle B B^\dagger \rangle \sigma^z 
\end{align}
These are related by the similarity transformation,
\begin{align}
    C=R^{-1}C_BR.
\end{align}

The ground state of a bosonic insulator with the Hamiltonian in \cref{eq:H_with_psi} is the state annihilated by all of the Bogoliubov mode annihilation operators ($b_-$ and $b_+$).
It is straightforward to show that the ground state correlation matrix ${C^g_{B}}$ is a diagonal projector onto the negative mode space, and hence $C^g$ is also a projector onto the negative mode space: 
\begin{align}
     {C^g_B} &= \langle BB^\dagger\rangle\sigma_z = \begin{bmatrix}
        \langle b_- b_-^\dagger \rangle & -\langle b_- b_+ \rangle \\
        \langle b_+^\dagger b_- \rangle & -\langle b_+^\dagger b_+ \rangle
    \end{bmatrix} = \begin{bmatrix}
        \mathbb{1}_{N\times N} & 0 \\0 &0
    \end{bmatrix} \nonumber\\
    C^g&=R^{-1}{C^g_B} R =\sum_{n} P_{-n}
    \label{eq:C_project_to_negative_space}
\end{align}
The projector $P$ in \cref{eq:alpha_cs,eq:alpha_G} is the ground state correlator of a lattice insulator in the momentum space (see \cref{eq:projector_p_from_c}). 

While it is evident that the ground state correlation matrix is a projector from the above relations, it can be shown that a more general set of correlation matrices are projectors; correlation matrices $C$ of states with zero charge obey (see \cref{sec:proof-of-c2c-for-zero-charge-states})
\begin{equation}
    C^2=C.
    \label{eq:c^2=c}
\end{equation}

\begin{figure*}[t]
    \centering
    \includegraphics[width=\linewidth]{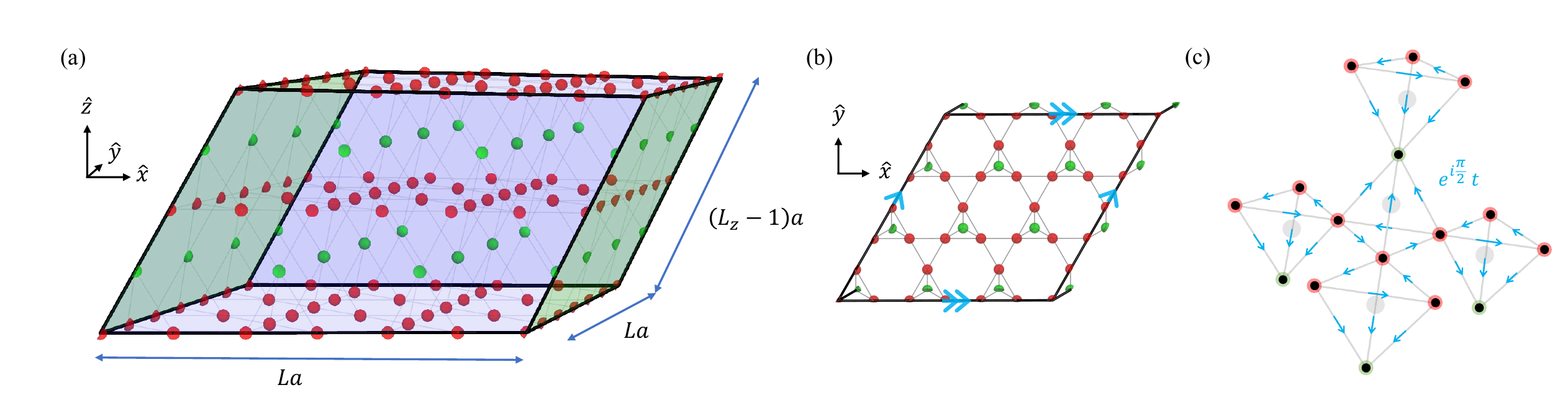}
    \caption{%
    The pyrochlore lattice can be understood as (a) alternating (111)-planes of Kagom\'e (red) and triangular (green) lattice or 
    (c) a system of corner-sharing tetrahedra. 
    The unit cell can be chosen as the four sites of an up-pointing tetrahedron ($\frac{a}{2}$ is the side length of the tetrahedron).
    In our model for the magnetoelectric effect, the red (green) sites have a positive (negative) potential. 
    We define a Cartesian coordinate system with the z-axis in the [111]-direction and the $x,y$-axes as shown.
    (a) 
    For real space calculations, we apply a magnetic field $B\hat{z}$ to a parallelepiped computational volume with open boundaries at the top and bottom Kagom\'e planes and periodic boundaries in the other two directions (the left and right faces and the front and back faces are glued).
    These boundary conditions preserve a global $\pi$-rotation so that there will be no zero-field polarization $P^z$ in the finite-size system. 
    The figure shown has $L=3$ and $L_z=3$. 
    (b) The top-down view of one layer of pyrochlore unit cells. 
    (c) A single unit cell in the lattice (the four sites of the central upward-facing tetrahedron) connected to sites from neighboring unit cells. The arrows indicate a gauge choice for the elements of the spring matrix which gives a system with background magnetic monopoles in all tetrahedra. }
    \label{fig:pyro_lattice}
\end{figure*}

\begin{figure*}
    \centering
    \includegraphics[width=\linewidth]{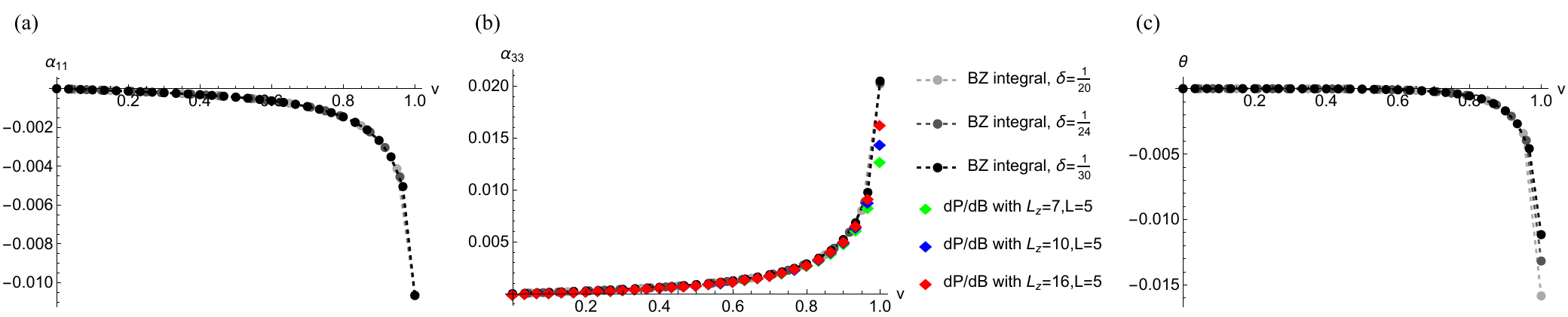}
    \caption{ Magnetoelectric response of the system of monopoles in a pyrochlore lattice with a staggered potential.  The rotational symmetries of the model ensures that $\alpha_{ij}$ is diagonal and $\alpha_{11}=\alpha_{22}$. The plots show  (a) $\alpha_{11}$, (b) $\alpha_{33}$ and (c) $\theta= 4\pi^2 \alpha_\theta$  as a function of the charge staggering strength $v$ with the other parameters chosen to be $\lambda=4.4642,t=1$ and $m=1$.   The grey and black curves are obtained by computing the integrals in \cref{eq:alpha_cs,eq:alpha_G} using the quadrature rule with a linear discretization size $\delta$. The colored points are obtained by measuring the polarization of different finite-sized pyrochlore lattices periodic in the $xy$-plane with $L\times L$ unit cells and open in the $z$-direction with $L_z$ unit cells, with a magnetic field applied along the $z$-direction (see \cref{fig:pyro_lattice}). Due to commensurability issues of the lattice vectors and the periodic planes, we restrict our finite-size computations to the response in the $z$-direction. Note that the parameters are chosen such that a large magnetoelectric effect can be seen. The gap decreases as $v$ goes to 1 for these parameters.  }
    \label{fig:alpha_pyro_monopoles}
\end{figure*}

The correlation matrix $C(t)$ of a time-evolved state satisfies the relation (see \cref{sec:proof-of-dot-cich-for-a-state-being-time-evolved-by-h})
\begin{align}
    i\dot C(t) = [h,C(t)].
    \label{eq:time_ev_C}
\end{align}

The expectation value of any quadratic operator, expressed as $\mathcal{M}= \psi^\dagger \sigma^y M \psi$, in the state whose correlation matrix is $C$ is given by 
\begin{align}
    \langle \mathcal{M}\rangle = \text{Tr} \ M(C-\mathbb 1).
    \label{eq:expectation_value_general}
\end{align}
Using the above expression, we can show that the total current density in a lattice is (refer \cref{sec:charge-and-current-operators})
\begin{align}
\langle {\bm{J}}_T \rangle= \frac{i}{\Omega} \text{Tr}\ C[{\bm{\mathcal{R}}},h] ,
\label{eq:current_bosonic}
\end{align}
where, $\Omega$ is the volume of a unit cell and $\bm{\mathcal{R}}$ is a position operator that specifies the position of the complex scalar in the bosonic lattice (refer to \cref{eq:postion_operator}).

\section{Numerical Demonstration for a Lattice model}\label{sec:numerics}

In this section, we compute $\alpha$ for a non-trivial pyrochlore lattice model using two independent methods: the $k$-space integration of \cref{eq:alpha_cs,eq:alpha_G} and finite size diagonalization in real space. 
The model that we present here does not apply to any physical system that we are aware of. Nonetheless, it serves as an example of a lattice bosonic insulator that shows a magnetoelectric response and allows us to verify our expression for $\alpha$.

A simple lattice bosonic insulator that shows a magnetoelectric response can be obtained by considering a pyrochlore lattice, with monopoles in all tetrahedra and a staggered potential in alternating planes. 
We consider the Hamiltonian in \cref{eq:hamiltonian_phi_pi}, with the sum on $r$ now going over all the pyrochlore lattice points, and the charge potential $V_r$ to be $-1$ for the points in the Kagom\'e planes perpendicular to the $z$-direction and $+1$ for the remaining points in the triangular lattice planes. The hopping matrix $K$ is set so every face of every tetrahedron has an outward flux of $\pi/2$, i.e., every tetrahedron has a magnetic monopole in it. 
The presence of monopoles breaks $\mathcal{P}$, $T$, and all the mirror symmetries, while the staggered charge potential breaks CT.

For finite-size lattice computations, we consider periodic boundary conditions in the $xy$-plane with $L$ unit cells in each direction and an open boundary with $L_z$ unit cells along the $z$-direction (see \cref{fig:pyro_lattice}). 
For this computation, one needs to make sure that the edges of the open boundary are both the same kind of planes (we consider Kagom\'e planes), to ensure that the system does not have finite polarization at zero external field.
The external magnetic field enters the Hamiltonian only via minimal coupling in the spring matrix elements:
\begin{align}
    K_{\bm{rr}'}(B)= K_{\bm{rr}'} \exp \left( i \int_{\bm{r}}^{\bm{r}'} \bm{A}(\bm{r}).d\bm{r}\right),
\end{align}
where, the vector potential $\bm{A}(\bm{r})$ is chosen such that it is periodic in the $xy$-plane and its curl is uniform i.e. $\nabla \times \bm{A} = B \hat z$. 
We diagonalize the Hamiltonian of this system subjected to small applied uniform external magnetic fields and obtain the charge distribution in the ground state (from \cref{eq:ground_state_charge}). 
This allows us to compute polarization and estimate the magnetoelectric response along the open direction.
The numerical values of $\alpha$ estimated in this system along with the values of $\alpha$ obtained from the numerical $\bm{k}$-space integration of \cref{eq:alpha_G,eq:alpha_cs} is shown in \cref{fig:alpha_pyro_monopoles}. 

\section{Outlook}\label{sec:outlook}

In this article, we discuss the ingredients required to construct quadratic microscopic models that exhibit magnetoelectric response in bosonic insulators. We derive an expression for the magnetoelectric response coefficient and numerically verify the expression by calculating the magnetoelectric tensor for the models we present. 
When compared to similar simple models of magnetoelectric fermionic systems, the bosonic systems have the added complexity of having to explicitly break $CT$ symmetry.

The derived expression allows for the calculation of the magnetoelectric response of bosonic insulators.
The primary example is the Mott insulating phase of ultracold bosonic atoms with lattice potentials designed to appropriately break inversion and time reversal symmetries. 
In the Mott insulating phase, the low-energy Hamiltonian conserves the total number of bosons in the system. 
Expanding around the ground state can give an effective quadratic $U(1)$ symmetric bosonic Hamiltonian with the same low energy spectrum as the Mott insulator. 
Our formalism can be applied to the effective quadratic Hamiltonian to estimate the magnetoelectric response of the system.
In this setup, the magnetoelectric response naturally shows up in two different probes.
Uniform synthetic magnetic fields\cite{goldmanLightinducedGaugeFields2014,grossQuantumSimulationsUltracold2017,schaferToolsQuantumSimulation2020} induce polarization, in which the bosons shift in the direction of the applied field.
Alternatively, tilting the lattice potential corresponds to a synthetic electric field which induces magnetization -- that is, microscopic loop currents of the underlying bosons.

A more exotic example is provided by the Coulomb quantum spin liquid \cite{udagawaSpinIce2021,henleyCoulombPhaseFrustrated2010,savaryCoulombicQuantumLiquids2012,savaryQuantumSpinIce2016} where coexistence with $\mathcal{P}$ and $T$ breaking orders can lead to axion electrodynamics \cite{paceDynamicalAxionsQuantum2023}. 
In these systems, a gauge mean field (gMFT) approximation leads to effective rotor models that describe the gapped bosonic spinon excitations. The rotors can be further approximated by quadratic bosons.
The methods developed here can be used to calculate the effective axion coupling by working with the effective bosonic Hamiltonian. 
Unlike the ultracold Mott insulator, here, the magnetoelectric effect is coupled to a true dynamical gauge field and thus can be probed through the emergent electrodynamic response.
We leave these avenues of study for future work.

\begin{acknowledgments}
The authors are grateful to
S. Pace, M. Bellitti, D. Long, Y.M. Lu, J. Moore, 
C. Chamon, A. Vishwanath,  and D. Arovas for stimulating discussions.
C.R.L. acknowledges support from the  National Science Foundation through Grant No. PHY-1752727.
\end{acknowledgments}

\appendix

\section{Hamiltonian of the Toy Models}
\label{sec:toy_model_hamiltonian}

\begin{figure}[b]
    \centering
    \includegraphics[width=.5\linewidth]{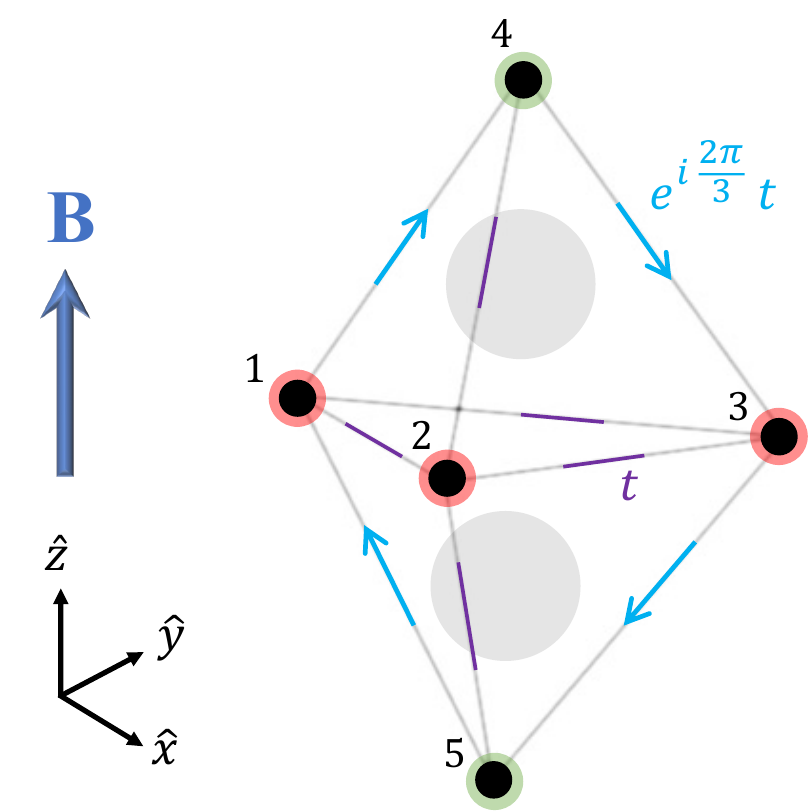}
    \caption{The bosonic bipyramid toy model with each site numbered to indicate the chosen basis. The magnetic field is applied along the $z$ direction. The arrows indicate a particular gauge choice for the spring matrix that gives a system with monopoles in both tetrahedra of the bipyramid.}
    \label{fig:annotated_bipyramid}
\end{figure}

All the toy models we study in this article have the Hamiltonian of the form in \cref{eq:simple_hamiltonian_pi_phi}. In this section, we explicitly specify the matrix form of the Hamiltonian for the bipyramid model. We also include the momentum space Hamiltonian of the pyrochlore lattice model.

\subsection{Bipyramid with monopoles}

We choose a position space basis for the bipyramid as specified in \cref{fig:annotated_bipyramid}. We set the length scale in the system by choosing each side of the bipyramid to be of length 1. We choose the $x$-axis to point from the first site to the second and the $z$-axis to point up, parallel to the three-fold symmetry axis. The Hamiltonian of this system under finite external magnetic field, $B\hat{z}$, is given by \cref{eq:simple_hamiltonian_pi_phi}, with the charge potential matrix, $V=\text{diag}(-1,-1,-1,1,1)$, and the spring matrix
\begin{align}
K(B)=\left(
\begin{array}{ccccc}
 0 & 1 & e^{-i \frac{\sqrt{3} B}{8} } & \omega^2 e^{ -\frac{i B}{8 \sqrt{3}}} & \omega e^{ -\frac{i B}{8 \sqrt{3}}} \\
 1 & 0 & e^{-i\frac{3  \sqrt{3} B}{8} } & e^{-i \frac{\sqrt{3} B}{8} } & e^{-i \frac{\sqrt{3} B}{8} } \\
 e^{i \frac{\sqrt{3} B}{8} } & e^{i\frac{3\sqrt{3} B}{8}  } & 0 & \omega e^{ \frac{i B}{2 \sqrt{3}}} & \omega^2 e^{ \frac{i B}{2 \sqrt{3}}} \\
 \omega e^{ \frac{i B}{8 \sqrt{3}}} & e^{i \frac{\sqrt{3} B}{8} } & \omega^2 e^{ -\frac{i B}{2 \sqrt{3}}} & 0 & 0 \\
 \omega^2 e^{ \frac{i B}{8 \sqrt{3}}} & e^{i \frac{\sqrt{3} B}{8} } & \omega e^{ -\frac{i B}{2 \sqrt{3}}} & 0 & 0 \\
\end{array}
\right),
\end{align}
where $\omega=e^{i\frac{2\pi}{3}}$. Under a finite external electric field, $\varepsilon\hat{z}$, the Hamiltonian has the charge potential matrix $V(\varepsilon)=\text{diag}\left(-1,-1,-1,1-\sqrt{\frac{2}{3}}\frac{\varepsilon}{v},1+\sqrt{\frac{2}{3}}\frac{\varepsilon}{v}\right)$ and spring matrix $K(B=0)$.

\subsection{Pyrochlore with monopoles}

Our lattice toy model (see \cref{sec:numerics} and \cref{fig:pyro_lattice}) has a four-site unit cell. We choose our unit cell to be the sites on an upward-facing tetrahedron, with our basis chosen so the apex of the tetrahedron is the first site in the unit cell. Under the absence of external fields, the Hamiltonian of this system in momentum space is given by:
\begin{align}
    \mathcal{H} = 
    \sum_{\bm{k}}\begin{pmatrix} \Pi^\dagger_{\bm{k}}&\Phi^\dagger_{\bm{k}} \end{pmatrix}
    \begin{pmatrix} m^{-1} \mathbb{1} & i v V_{\bm{k}} \\
    -iv V_{\bm{k}} & \lambda \mathbb{1} + t\  K_{\bm{k}} \end{pmatrix}
    \begin{pmatrix} \Pi_{\bm{k}} \\
    \Phi_{\bm{k}} \end{pmatrix}
\end{align}
where, $\mathbb{1}$ is the $4\times 4$ identity matrix, $V_k=\text{diag}(1,-1,-1,-1)$ is the diagonal charge matrix,

\begin{widetext}
\begin{align}
K_{\bm{k}}=2\left(
\begin{array}{cccc}
 0  & i \cos \left(\frac{1}{2} \bm{k}.\left(\bm{e}_1-\bm{e}_2\right)\right) & i \cos \left(\frac{1}{2} \bm{k}.\left(\bm{e}_1-\bm{e}_3\right)\right) & i \cos \left(\frac{1}{2} \bm{k}.\left(\bm{e}_1-\bm{e}_4\right)\right) \\
 -i \cos \left(\frac{1}{2} \bm{k}.\left(\bm{e}_1-\bm{e}_2\right)\right) & 0  & - \sin \left(\frac{1}{2} \bm{k}.\left(\bm{e}_2-\bm{e}_3\right)\right) & - \sin \left(\frac{1}{2} \bm{k}.\left(\bm{e}_4-\bm{e}_2\right)\right) \\
 -i \cos \left(\frac{1}{2} \bm{k}.\left(\bm{e}_1-\bm{e}_3\right)\right) & - \sin \left(\frac{1}{2} \bm{k}.\left(\bm{e}_2-\bm{e}_3\right)\right) & 0  & - \sin \left(\frac{1}{2} \bm{k}.\left(\bm{e}_3-\bm{e}_4\right)\right) \\
 -i \cos \left(\frac{1}{2} \bm{k}.\left(\bm{e}_1-\bm{e}_4\right)\right) & - \sin \left(\frac{1}{2} \bm{k}.\left(\bm{e}_4-\bm{e}_2\right)\right) & - \sin \left(\frac{1}{2} \bm{k}.\left(\bm{e}_3-\bm{e}_4\right)\right) & 0  \\
\end{array}\right)
\end{align}
\end{widetext}
is the spring matrix, and the vectors, $\bm{e}_i$, point from the center of a unit cell to the lattice points. We choose these vectors to be:
\begin{align}
\bm{e}_1&= \{0,0,1\} \nonumber \\ 
\bm{e}_2&= \left\{0,-\frac{2 \sqrt{2}}{3} ,-\frac{1}{3}\right\}\nonumber \\ 
\bm{e}_3&=\left\{\sqrt{\frac{2}{3}},\frac{\sqrt{2}}{3},-\frac{1}{3}\right\}\nonumber \\ 
\bm{e}_4&=\left\{-\sqrt{\frac{2}{3}},\frac{\sqrt{2}}{3},-\frac{1}{3}\right\}
\end{align}

\section{Properties of the correlation matrix}\label{sec:properties_of_the_correlation_matrix}

\subsection{Proof of $C^2=C$ for states with zero charge}\label{sec:proof-of-c2c-for-zero-charge-states}

For a considered state, we define the matrix 
\begin{align}
    \mathcal{C}=\langle\psi \psi^\dagger\rangle,
    \label{eq:definition_curly_C}
\end{align}
so the correlation matrix in the same state is given by $C=\mathcal{C} \sigma^y$.
Since the operator $\psi^\dagger$ is a charge-lowering operator, for the correlations in a zero-charge state, we get
\begin{align}
\mathcal{C}&=\langle\psi \psi^\dagger \rangle = \langle\psi\  (-Q)\ \psi^\dagger \rangle \nonumber \\
&= \langle\psi\ ( \psi^\dagger \sigma^y \psi+1) \ \psi^\dagger \rangle\nonumber\\
&= \langle\psi_i\  \psi_k^\dagger {\sigma^y}_{kl} \psi_l \ \psi^\dagger_j \rangle+\langle \psi_i\psi_j^\dagger\rangle \nonumber\\
&= \langle\psi_i\  \psi_k^\dagger \rangle {\sigma^y}_{kl} \langle \psi_l \ \psi^\dagger_j \rangle + 
\langle\psi_i\ \psi^\dagger_j \rangle \langle (\psi_k^\dagger {\sigma^y}_{kl} \psi_l +1)\rangle \nonumber\\
&=\mathcal{C}\sigma^y \mathcal{C} - \mathcal{C}\ \langle Q\rangle
\end{align}
We have only specified the Nambu indices in the above equations. We used Wick's theorem for the simplification above. The second term in the last equation is zero if the charge of the state is zero. This leaves us with $\mathcal{C} = \mathcal{C} \sigma^y \mathcal{C}$, which leads to $C=C^2$.

\subsection{Proof of $\dot C=i[C,h]$ }\label{sec:proof-of-dot-cich-for-a-state-being-time-evolved-by-h}

In a system with Hamiltonian $\mathcal{H}= \psi^\dagger \sigma^y h \ \psi$, the equations of motion are:
\begin{align}
i \dot \psi = h\psi \quad \text{and} \quad i\dot\psi^\dagger = -\psi^\dagger \sigma^y  h \ \sigma^y,    
\end{align}
Taking the time derivative of the product and then the state expectation value,
\begin{align}
i \partial_t(\psi \psi^\dagger) &= h \psi\psi^\dagger - \psi \psi^\dagger \sigma^y  h \ \sigma^y \nonumber \\
i \dot{\mathcal{C}}&= h \mathcal{C}- \mathcal{C} \sigma^y h \ \sigma^y
\end{align}
Multiplying the above equation with $\sigma^y$ from the right, we get $$i\dot C=[h,C].$$

\section{Derivation of the magnetoelectric tensor}\label{sec:derivation-of-the-magnetoelectric-tensor}

We consider a time-dependent Hamiltonian whose magnetoelectric response varies with time.  
When this system is subjected to a constant magnetic field, it gains polarization over time, and the current leading to the polarization is a bulk current (see \cref{fig:bulk_current}). We analytically calculate the current and integrate it to obtain the polarization and extract the linear response coefficient $\alpha$. 
This derivation is a bosonic version of the derivation by Essin \textit{et al.}~\cite{essinOrbitalMagnetoelectricCoupling2010}.
We start with the mathematical descriptions of magnetic translational symmetry (MTS) and current in bosonic systems and then go on to the derivation.

\subsection{Magnetic Translational Symmetry}\label{sec:magnetic-translational-symmetry-mts}
Consider a translationally symmetric system in a uniform magnetic field. The Hamiltonian in the position basis is given by
\begin{align}
\mathcal{H}= \sum_{i,j,\bm{r},\bm{r}',\alpha,\alpha'} \psi^\dagger_{i\bm{r}\alpha}\  (\sigma^y \ h)_{i\bm{r}\alpha,j\bm{r}'\alpha'} \ \psi_{j\bm{r}'\alpha'}.
\end{align}
where, for $\psi_{i\bm{r}\alpha}$, $i$ is the Nambu index that specifies if $\Phi$ or $\Pi$ is chosen, $\bm{r}$ specifies the position of the unit cell, and $\alpha$ is the unit cell index which specifies the position to be $\bm{r}+\delta \bm{r}_\alpha$.
Despite the presence of translational symmetry, the Hamiltonian (and the dynamical matrix $h$) will not have the same symmetry, since the vector potential corresponding to a uniform magnetic field cannot be chosen to be translationally symmetric. However, $h$ can be broken down into a translationally symmetric term times a phase, which depends on the choice of gauge for the vector potential.

We choose the symmetric gauge for the vector potential,
$\bf {\mathcal A}(r) = \frac{1}{2} \bf B\times \bf r$. Then, $h$ is said to have MTS if it satisfies the relation
\begin{align}
h_{i({\bm{r}}+{\bm{r}}_o)\alpha,j({\bm{r}}'+{\bm{r}}_o)\alpha'}= e^{\frac{i}{2} {\bm{B}}\cdot ({\bm{r}}_o \times ({\bm{r}}-{\bm{r'}}))} \ h_{i{\bm{r}}\alpha,i \bm{r'}\alpha'} .
\end{align}
If $h$ has MTS, then it can be expressed as
\begin{align} 
h_{i\bm{r}\alpha,j\bm{r}'\alpha'} = \bar{h}_{i\bm{r}\alpha,j\bm{r}'\alpha'} \ e^{-\frac{i}{2} {\bm{B}}\cdot ({\bm{r}} \times {\bm{r}}')}
\end{align}
where, $\bar{h}$ is translationally symmetric. Further,
\begin{align}
\bar h = h_o +h'(B),
\end{align}
where, $h_o$ is the dynamical matrix of the system when $B=0$ and $h'(B)$ is the dependence of the dynamical matrix on $B$ that isn't accounted for by adding minimal coupling to the hopping matrix elements.

If the Hamiltonian (or dynamical matrix) of a system has MTS, then the correlation matrix corresponding to the ground state of the same Hamiltonian also has MTS and
\begin{align}
C_{i{\bm{r}\alpha},j{\bm{r}}'\alpha'} &= \bar{C}_{i{\bm{r}}\alpha,j{\bm{r}}'\alpha'} e^{- \frac{i}{2} {\bm{B}}\cdot({\bm{r}}\times {\bm{r}}')} \\
\bar{C}&=C_o +C'
\end{align}
where, $\bar{C}$ is translationally symmetric, $C_o$ is independent of ${\bm{B}}$ and $C'$ encodes all the dependence of $\bar C$ on ${\bm{B}}$.

\subsection{Charge and Current 
 operators}\label{sec:charge-and-current-operators}
The total charge of the system is given by \cref{eq:total_charge}.
The local charge at a point ${\bm{r}}+\delta \bm{r}_\alpha$ is
\begin{align}
Q_{{\bm{r}}\alpha} = -\sum_{ij}\psi_{i\bm{r}\alpha}^\dagger \sigma^y_{ij} \psi_{j\bm{r}\alpha}-1    
\end{align} 
The expectation of local charge in a state with correlation matrix $C$ (using \cref{eq:expectation_value_general}) is given by:
\begin{align}
    \langle Q_{\bm{r}\alpha}\rangle= 1-C_{1\bm{r}\alpha}-C_{2\bm{r}\alpha}
    \label{eq:ground_state_charge}
\end{align}
In the equations that follow, we suppress the Nambu indices when they are unnecessary. 
The local current can be found by using the continuity equation $\text{div}\left.(\mathcal{\bm{J}})\right|_{\bm{r}+\delta \bm{r}_\alpha} = \partial_t Q_{\bm{r}\alpha}$  to be
\begin{align}
\mathcal{\bm{J}}_{{\bm{r}}\alpha,{\bm{r}}'\alpha'} = i \bigg( & \psi_{\bm{r}\alpha}^\dagger \left(\sigma^y h\right)_{{\bm{r}}\alpha,{\bm{r}}'\alpha'}\  \psi_{{\bm{r}}'\alpha'} \nonumber\\
&-\psi_{{\bm{r}}'\alpha'}^\dagger \left(\sigma^y h\right)_{{\bm{r}}'\alpha',{\bm{r}}\alpha} \ \psi_{{\bm{r}}\alpha}\bigg).
\end{align}
The local current density vector is given by
\begin{align}
{\bm{J}}_{\bm{r}\alpha} = \frac{1}{\Omega} \sum_{{\bm{r}}',\alpha'} ({\bm{r}}+\delta \bm{r}_\alpha-{\bm{r}}'-\delta \bm{r}_{\alpha'}) \mathcal{\bm{J}}_{{\bm{r}\alpha}{\bm{r}}'\alpha'}
\end{align}
where, $\Omega$ is the volume of a unit cell. The total current density is ${\bm{J}}_T = \sum_{\bm{r},\alpha} {\bm{J}}_{\bm{r}\alpha}$

We now define a position operator in this space as
\begin{align}
{\bm{\mathcal{R}}}_{i{\bm{r}}\alpha,j{\bm{r}}'\alpha'} = ({\bm{r}} + \delta \bm{r}_\alpha) \delta_{{\bm{r}}{\bm{r}}'} \delta_{\alpha,\alpha'} \delta_{ij}.
\label{eq:postion_operator}
\end{align}
With the above definition of ${\bm{\mathcal{R}}}$, we can now express the total current density of a state with correlator $C$ as
\begin{align}
{\bm{J}} = \langle {\bm{J}}_T \rangle= \frac{i}{\Omega} \text{Tr}\ C[h,{\bm{\mathcal{R}}}] .
\end{align}
Using \cref{eq:c^2=c} the total current density can be expressed as
\begin{align}
{\bm{J}} &= \frac{i}{\Omega} \text{Tr}\ [ C,[C,{\bm{\mathcal{R}}}]][C,h] 
\end{align}
For a state being time evolved by the dynamical matrix $h$, using \cref{eq:time_ev_C}, the total current density can be expressed as
\begin{align}
{\bm{J}} &= \frac{1}{\Omega} \text{Tr}\ [ C,[C,{\bm{\mathcal{R}}}]]\ \dot{C}. 
\label{eq:jindotc}
\end{align}

\subsection{Hamiltonian and state correlators}

We consider a time-dependent Hamiltonian whose magnetoelectric response varies with time.  
Consider a $U(1)$ conserving system of non-interacting bosons in a uniform magnetic field ${\bm{B}}$, with the Hamiltonian
$\mathcal{H}(\beta,{\bm{B}}) = \psi^\dagger \sigma^y h \ \psi$ containing a
 time dependant parameter $\beta(t)$, which is such that the magnetoelectric polarizibility $\alpha$ vanishes when $\beta =0$. 
 For example, $\beta$ could be the parameter that accompanies a term that breaks $\mathcal{P}$ (parity) or $T$ (time-reversal) or both symmetries.

Let $C^g$ and $C_o$ be the correlators corresponding to the ground state of $\mathcal{H}(\beta,{\bm{B}})$ and $\mathcal{H}(\beta, {\bm{B}}=0)$ respectively.
We imagine a situation where the system at time $t=0$ is initialized in the ground state of $\mathcal{H}(\beta=0,{\bm{B}})$ and is adiabatically evolved with the Hamiltonian $\mathcal{H}(\beta(t),{\bm{B}})$, where $\beta(0)=0$. The correlator corresponding to the state at time $t$ is $C(t)$.

The derivation can be broken down into two steps: First, we find the perturbative corrections to $C(t)$ in $\bm{B}$ using MTS, and second, we integrate the perturbative expansion of the current. In the rest of the derivation, we suppress the Nambu index and the unit cell index when they are not important.

\subsection{Perturbative expansion in $\mathbf{B}$ using MTS}\label{sec:perturbative-expansion-in-b-using-mts}

Since the system has MTS (see \cref{sec:magnetic-translational-symmetry-mts}) , we have
\begin{align}
C^g_{{\bm{r}}_1{\bm{r}}_2} &= \bar{C^g}_{{\bm{r}}_1{\bm{r}}_2} e^{-\frac{i}{2} {\bm{B}}\cdot({\bm{r}}_1 \times {\bm{r}}_2)}\nonumber\\
\bar{C^g}&= C_o+C'
\label{eq:mtsonc}
\end{align}
Using a Fourier transformation
\begin{align}
    \psi_{\bm{r}\alpha} =  \int_{\text{BZ}} \frac{d^3{k}}{(2\pi)^3}  e^{i{\bm{k}}\cdot\left({\bm{r}}+\delta \bm{r}_\alpha\right)} \psi_{\bm{k}\alpha},
\end{align}
the ground state correlator at zero magnetic field $C_o$ is simply given by
\begin{align}\label{eq:projector_p_from_c}
{C_o}_{{\bm{r}}_1{\bm{r}}_2} &= \int_{\text{BZ}} \frac{d^3{k}}{(2\pi)^3} e^{i{\bm{k}}\cdot{\bm{r}}_1} P_{\bm{k}} e^{-i{\bm{k}}\cdot{\bm{r}}_2}, \\
\text{with} \; P_{\bm{k}} &= \sum_{n=1}^N \mathcal{w}_{{\bm{k}}n}\mathcal{v}^\dagger_{{\bm{k}}n}, \nonumber
\end{align}
where, $\mathcal{w}_{{\bm{k}}n}$ and $\mathcal{v}^\dagger_{{\bm{k}}n}$ are the right and left eigenvectors of  ${h_o}_{\bm{k}}$, the momentum space dynamical matrix of the Hamiltonian $\mathcal{H}(\beta,{\bm{B}}=0)$. 
$C'$ is the ${\bm{B}}$ dependent component of $\bar{C^g}$ and we find $C'$ to first order in ${\bm{B}}$ by using some simple relations and MTS.

The ground state of the system has zero total charge, and hence we have (see \cref{sec:proof-of-c2c-for-zero-charge-states}):
\begin{align}
C^g =(C^g)^2 
\end{align}
Expanding the above expression in the position basis and using MTS, we get
\begin{align}
\bar{C^g}_{{\bm{r}}_1{\bm{r}}_3}= \sum_{{\bm{r}}_2} \bar{C^g}_{{\bm{r}}_1{\bm{r}}_2} \bar{C^g}_{{\bm{r}}_2{\bm{r}}_3} e^{-\frac{i}{2}{\bm{B}}\cdot\left({\bm{r}}_1\times{\bm{r}}_2+{\bm{r}}_2\times{\bm{r}}_3+{\bm{r}}_3\times{\bm{r}}_1\right)}
\end{align}
Expanding the above equation to first order in ${\bm{B}}$ and using the relation $\left({\bm{r}}_1\times{\bm{r}}_2+{\bm{r}}_2\times{\bm{r}}_3+{\bm{r}}_3\times{\bm{r}}_1\right) = ({\bm{r}}_2 -{\bm{r}}_1)\times ({\bm{r}}_3-{\bm{r}}_2)$, we get
\begin{align}
(1-C_o) C' (1-C_o) - C_o C' C_o = -\frac{i}{2} {\bm{B}}\cdot[C_o,{\bm{\mathcal{R}}}]\times [C_o,{\bm{\mathcal{R}}}]
\end{align}
Recall that $C_o$ is a projector, and hence, the above equation allows us to find all the components of $C'$ projected onto the same negative/positive mode.

To find the remaining `off-diagonal' components of $C'$, note that the matrices $h$ and $C^g$ have the same eigenstates, and hence we have
\begin{align}
[h,C^g]=0.
\end{align}
Rewriting the above equation in the position basis, using MTS (see \cref{sec:magnetic-translational-symmetry-mts}) and expanding to linear order in ${\bm{B}}$ gives:
\begin{align}
[C',h_o] = & \frac{i}{2} {\bm{B}}\cdot\left([ C_o ,{\bm{\mathcal{R}}}] \times[h_o,{\bm{\mathcal{R}}}] - [h_o,{\bm{\mathcal{R}}}]\times [C_o,{\bm{\mathcal{R}}}] \right) 
\nonumber\\ &- [C_o,h'] 
\end{align}
Using \cref{eq:projector_p_from_c,eq:projector_nth_mode}, we get
\begin{align} 
P_{-n} C' P_{+m} = &\frac{ i \ B^j \epsilon_{jab} \ P_{-n} \{\partial^a {h_o}_, \partial^b P \}P_{+m}}{E_{-n} +E_{+m}} \nonumber\\
 &+ \frac{  P_{-n}h' P_{+m}}{E_{-n} +E_{+m}}.
\label{eq:crossgapelements}
\end{align}
In the above equation, momentum labels have been dropped for brevity.

\subsection{Expanding Current in ${\mathbf{B}}$}\label{sec:expanding-current-in-b}
With the thought experiment considered, the adiabatic time evolution of the state would be given by
\begin{align}
 i \dot C(t) = [ h(t), C(t)].
\end{align}
The total current density at time $t$ is given by
\begin{align}
{\bm{J}} &= \frac{1}{\Omega} \text{Tr}\ [ C,[C,{\bm{\mathcal{R}}}]]\ \dot{C} 
\end{align}
Using the adiabatic approximation $C\approx C^g$ in the above equation and expanding in position space,
\begin{align}
{\bm{J}} = \frac{1}{\Omega} \sum_{{\bm{r}}_1,{\bm{r}}_2,{\bm{r}}_3} ({\bm{r}}_1 -2{\bm{r}}_2+{\bm{r}}_3)\   C^g_{{\bm{r}}_1{\bm{r}}_2}C^g_{{\bm{r}}_2{\bm{r}}_3} \dot{C^g}_{{\bm{r}}_3{\bm{r}}_1} 
\end{align}
Using \cref{eq:mtsonc} and expanding linear in ${\bm{B}}$, we get

\begin{widetext}
\begin{dmath}
    J = \frac{1}{\Omega}\sum_{{\bm{r}}_1,{\bm{r}}_2,{\bm{r}}_3} ({\bm{r}}_1 -2{\bm{r}}_2+{\bm{r}}_3)\left(   {C_o}_{{\bm{r}}_1{\bm{r}}_2} {C_o}_{{\bm{r}}_2{\bm{r}}_3} \dot{C'}_{{\bm{r}}_3{\bm{r}}_1}
+ {C'}_{{\bm{r}}_1{\bm{r}}_2} {C_o}_{{\bm{r}}_2{\bm{r}}_3} \dot{C_o}_{{\bm{r}}_3{\bm{r}}_1}
+ {C_o}_{{\bm{r}}_1{\bm{r}}_2} {C'}_{{\bm{r}}_2{\bm{r}}_3} \dot{C_o}_{{\bm{r}}_3{\bm{r}}_1}\\
 - \frac{i}{2}{\bm{B}}\cdot\left({\bm{r}}_1\times{\bm{r}}_2+{\bm{r}}_2\times{\bm{r}}_3+{\bm{r}}_3\times{\bm{r}}_1\right)\   {C_o}_{{\bm{r}}_1{\bm{r}}_2} {C_o}_{{\bm{r}}_2{\bm{r}}_3} \dot{C_o}_{{\bm{r}}_3{\bm{r}}_1} \right)
\end{dmath}
\end{widetext}
\begin{comment}
\begin{widetext}
\begin{align}
J = \frac{1}{\Omega}\sum_{{\bm{r}}_1,{\bm{r}}_2,{\bm{r}}_3} ({\bm{r}}_1 -2{\bm{r}}_2+{\bm{r}}_3)\ \left(  \begin{aligned} {C_o}_{{\bm{r}}_1{\bm{r}}_2} {C_o}_{{\bm{r}}_2{\bm{r}}_3} \dot{C'}_{{\bm{r}}_3{\bm{r}}_1}
+ {C'}_{{\bm{r}}_1{\bm{r}}_2} {C_o}_{{\bm{r}}_2{\bm{r}}_3} \dot{C_o}_{{\bm{r}}_3{\bm{r}}_1}
+ {C_o}_{{\bm{r}}_1{\bm{r}}_2} {C'}_{{\bm{r}}_2{\bm{r}}_3} \dot{C_o}_{{\bm{r}}_3{\bm{r}}_1}\\
 - \frac{i}{2}{\bm{B}}\cdot\left({\bm{r}}_1\times{\bm{r}}_2+{\bm{r}}_2\times{\bm{r}}_3+{\bm{r}}_3\times{\bm{r}}_1\right)\   {C_o}_{{\bm{r}}_1{\bm{r}}_2} {C_o}_{{\bm{r}}_2{\bm{r}}_3} \dot{C_o}_{{\bm{r}}_3{\bm{r}}_1} \end{aligned} \right)
\end{align}
\end{widetext}
\end{comment}
The first term above can be expanded in terms of a total derivative
\begin{align}
    {C_o}_{{\bm{r}}_1{\bm{r}}_2} {C_o}_{{\bm{r}}_2{\bm{r}}_3} \dot{C'}_{{\bm{r}}_3{\bm{r}}_1} &= \partial_t( {C_o}_{{\bm{r}}_1{\bm{r}}_2} {C_o}_{{\bm{r}}_2{\bm{r}}_3} {C'}_{{\bm{r}}_3{\bm{r}}_1})\nonumber\\
    - \dot{C_o}_{{\bm{r}}_1{\bm{r}}_2} {C_o}_{{\bm{r}}_2{\bm{r}}_3} & {C'}_{{\bm{r}}_3{\bm{r}}_1}
-{C_o}_{{\bm{r}}_1{\bm{r}}_2} \dot{C_o}_{{\bm{r}}_2{\bm{r}}_3} {C'}_{{\bm{r}}_3{\bm{r}}_1}.
\end{align}
The total derivative in the above equation can be rewritten as
\begin{align}
{\bm{J}}_{\text{G}} = \frac{1}{\Omega} \partial_t \left( \text{Tr}\ [C_o,{\bm{\mathcal{R}}}] [ C', C_o] \right)
\end{align}
Using the expression for the off-diagonal elements of $C'$ (see \cref{eq:crossgapelements}) in the above expression and using $J_{\text{G}}^i = \partial_t(\alpha_{\text{G}})^i _j B^j$, we get the cross gap contribution to the magnetoelectric polarizability, $\alpha_{\text{G}}$, in \cref{eq:alpha_G}

The rest of the terms in the equation for total current density can be rewritten as
\begin{align}
{\bm{J}}_{\text{CS}1} &= -\frac{3}{\Omega} \text{Tr} \ C' [\dot C_o , [ C_o, {\bm{\mathcal{R}}}]]\\
{\bm{J}}_{\text{CS}2} &= - \frac{i}{2} B^j \epsilon_{jab} \text{Tr}\ [C_o,{\bm{\mathcal{R}}}] [C_o,{\mathcal{R}}^a] [{\mathcal{R}}^b,\dot{C_o}] + \text{c.c.}
\end{align}
Taking the Fourier transform and simplifying, we get
\begin{comment}
\begin{align}
{\bm{J}}_{\text{CS}} &= {\bm{J}}_{\text{CS}1}+{\bm{J}}_{\text{CS}2} \nonumber\\
 &=\int_{\text{BZ}} \frac{d^3{k}}{(2\pi)^3} {\bm{B}}\  \text{Tr}\ P_{\bm{k}}&& \bigg(
 [\dot P_{\bm{k}}, \partial^x P_{\bm{k}}][\partial^y P_{\bm{k}},\partial^z P_{\bm{k}}] \nonumber\\
 &  &&+[\dot P_{\bm{k}}, \partial^y P_{\bm{k}}][\partial^z P_{\bm{k}},\partial^x P_{\bm{k}}] \nonumber\\
  &  && +[\dot P_{\bm{k}}, \partial^z P_{\bm{k}}][\partial^x P_{\bm{k}},\partial^y P_{\bm{k}}] \bigg)
\end{align}
\end{comment}
\begin{align}
{\bm{J}}_{\text{CS}} &= {\bm{J}}_{\text{CS}1}+{\bm{J}}_{\text{CS}2} \nonumber\\
 &=\int_{\text{BZ}} \frac{d^3{k}}{(2\pi)^3} {\bm{B}}\  \text{Tr}\ P_{\bm{k}} \bigg(
 [\dot P_{\bm{k}}, \partial^x P_{\bm{k}}][\partial^y P_{\bm{k}},\partial^z P_{\bm{k}}] \nonumber\\
 &  +[\dot P_{\bm{k}}, \partial^y P_{\bm{k}}][\partial^z P_{\bm{k}},\partial^x P_{\bm{k}}]+[\dot P_{\bm{k}}, \partial^z P_{\bm{k}}][\partial^x P_{\bm{k}},\partial^y P_{\bm{k}}] \bigg)
\end{align}

Using $F^{\mu\nu}= i P\left(\partial^\mu P \partial^\nu P - \partial^\nu P \partial^\mu P \right) P$, the above equation can be rewritten as
\begin{align}
{\bm{J}}_{\text{CS}}&= -\frac{{\bm{B}}}{8}\int_{\text{BZ}} \frac{d^3{k}}{(2\pi)^3} \epsilon_{\mu\nu\gamma\lambda} \ \text{Tr}\ F^{\mu\nu}F^{\gamma\lambda},
\end{align}
which, along with ${\bm{J}}_{\text{CS}}= \partial_t \alpha_{\text{CS}} \ {\bm{B}}$, gives us the expression for $\alpha_{\text{CS}}$ in \cref{eq:alpha_cs_chernform}.

\section{Group structure of the generalized Bogoliubov transformation}
\label{sec:group_structure_of_the_generalized_Bogoliubov_transformation}

The set of generalized Bogoliubov transformations $R$ defined in \cref{eq:h_diag_R} that diagonalize dynamical matrices $h$ don't form a group themselves. However, we can define transformations $S$ for every $R$ by 
\begin{align}
    R=S e^{-i \frac{\pi}{4} \sigma^x}.
\end{align}
With this, for the transformations $S$, we have the condition:
\begin{align}
    S\sigma^z S^\dagger=\sigma^z.
\end{align}
The transformations $S$ are part of the generalized unitary group $U(N,N)$. Note that $U(N,N) \cong G(2N)$, where $G$ is the conjugate symplectic group.
 
\section{Generality of formalism}\label{sec:generality_of_formalism}

The formalism introduced in \cref{sec:formalism} can be further generalized to any bosonic system with a $U(1)$ symmetry.
We can consider a system of bosons that have $N$ charge raising creation operators $a_{+i}^{\dagger}$ defined by $[Q,a_{+i}^{\dagger}]=a_{+i}^{\dagger}$ and $M$ charge lowering operators $a_{-i}^{\dagger}$ defined by $[Q,a_{-i}^{\dagger}]=-a_{-i}^{\dagger}$. The most general quadratic Hamiltonian conserving the charge $Q$ can be expressed in terms of $A=\left[\begin{smallmatrix}  a_-\\  a_+^\dagger \end{smallmatrix}\right]$, an  $N+M$ size column vector of operators, as
\begin{align}
\mathcal{H}&=A^{\dagger} o^z h \ A  \\ 
\text{with } o^z&= \begin{bmatrix}
 \mathbb 1_{M\times M} & 0 \\ 0 & -\mathbb{1}_{N\times N} \nonumber 
\end{bmatrix}
\end{align}
where, $o^z= [A,A^{\dagger}]$ is the matrix that specifies the commutation relations of the bosonic operators. This Hamiltonian can be diagonalized by the similarity transformation $S$, which gives the diagonal Bogoluibons \begin{align}
    B=SA=\left[\begin{smallmatrix}b_-\\  b_+^\dagger \end{smallmatrix}\right].
\end{align} These are operators that satisfy $[Q,b_{\pm i}^{\dagger}]=\pm b_{\pm i}^{\dagger}$ and $[H,b^{\dagger}_{\pm i}]=E_{\pm i} b^{\dagger}_{\pm i}$. 
The required transformation can be found by diagonalizing the matrix $h$ and ensuring that the new set of operators thus found still satisfies the bosonic commutation relations:
\begin{align}
h=S^{-1}\Lambda \ S \quad \text{and} \quad S\ o^zS^{\dagger}= o^z.
\label{eq:h_diag_S}
\end{align}
The systems considered in the main text are those that have $N=M$, where every pair of operators $a_{\pm i}$ can be thought of as charge increasing and decreasing modes of the local 2D oscillator.

\end{document}